\documentclass[sn-mathphys,Numbered]{sn-jnl}

\usepackage{graphicx}%
\usepackage{multirow}%
\usepackage{amsmath,amssymb,amsfonts}%
\usepackage{amsthm}%
\usepackage{mathrsfs}%
\usepackage[title]{appendix}%
\usepackage{xcolor}%
\usepackage{textcomp}%
\usepackage{manyfoot}%
\usepackage{booktabs}%
\usepackage{algorithm}%
\usepackage{algorithmicx}%
\usepackage{algpseudocode}%
\usepackage{listings}%

\begin{document}

\title[Article Title]{Direct observation of split-mode exciton-polaritons in a single MoS$_2$ nanotube
}


\author[1]{\fnm{A.I.} \sur{Galimov}}
\author[1]{\fnm{D.R.} \sur{Kazanov}}
\author[1]{\fnm{A.V.} \sur{Poshakinskiy}}
\author[1]{\fnm{M.V.} \sur{Rakhlin}}
\author[1]{\fnm{I.A.} \sur{Eliseyev}}
\author[1]{\fnm{A.A.} \sur{Toropov}}
\author[2]{\fnm{M.} \sur{Remskar}}
\author*[1]{\fnm{T.V.} \sur{Shubina}\email{shubina@beam.ioffe.ru}}

\affil*[1]{\orgname{Ioffe Institute}, \orgaddress{\city{St. Petersburg}, \postcode{194021}, \country{Russia}}}

\affil[2]{\orgdiv{Condensed Matter Physics Department}, \orgname{Jozef Stefan Institute}, \orgaddress{\street{Jamova cesta 39}, \city{Ljubljana}, \postcode{1000}, \country{Slovenia}}}


\abstract{
A single nanotube synthesized from a transition metal dichalcogenide (TMDC) exhibits strong exciton resonances and, in addition, can support optical whispering gallery modes. This combination is promising for observing exciton-polaritons without an external cavity. However, traditional energy-momentum-resolved detection methods are unsuitable for this tiny object. Instead, we propose to use split optical modes in a twisted nanotube with the flattened cross-section, where a gradually decreasing gap between the opposite walls leads to a change in mode energy, similar to the effect of the barrier width on the eigenenergies in the double-well potential. Using micro-reflectance spectroscopy, we investigated the rich pattern of polariton branches in single MoS$_2$ tubes with both variable and constant gaps. Observed Rabi splitting in the 40 -- 60 meV range is comparable to that for a MoS$_2$ monolayer in a microcavity. Our results, based on the polariton dispersion measurements and polariton dynamics analysis, present a single TMDC nanotube as a perfect polaritonic structure for nanophotonics.
}




\maketitle

\section*{Introduction}\label{intro}

Transition metal dichalcogenides (TMDCs), such as MoX$_2$ and WX$_2$ (X=S, Se), have high binding energy and giant oscillator strength of direct A and B excitons in both atomically thin and multilayer structures \cite{Evans1965, Wang2018}. This leads to a strong interaction between light and matter, resulting in the formation of an exciton-polariton (or polariton for short). The polariton, that is, a quantum superposition of an exciton and a photon mode, was observed in many semiconductor systems, being of practical importance for nanophotonics (threshold-less lasers \cite{Imamoglu1996}) and for quantum technologies (computing and simulators \cite{Ghosh2020, Kim2017}). Of greatest interest are polaritons in microcavities, where excitons interact with the strongly confined electromagnetic field \cite{Kavokin2008, Luo2023, Wei2023}, which increases the interaction energy and makes it greater than the broadening of the bare modes. This situation, called the strong coupling regime, manifests itself as an anticrossing of the upper and lower polariton branches in the dependence of the eigenenergies on the detuning between the exciton and cavity modes \cite{Weisbuch1992}. The Rabi splitting $\hbar \Omega_{\rm Rabi}=2g$ characterizes the frequency of oscillation between the photon and exciton states, where $g$ is a coupling strength.

A quasi-two-dimensional (2D) polariton is characteristic of a quantum well in a microcavity \cite{Kulakovskii2000}. Its extreme case is the polariton formed when a TMDC monolayer is placed in an external resonator \cite{Liu2015,Dufferwiel2015, Flatten2016, Zhang2018, Schneider2018, Lackner2021}.  All studies of 2D polaritons have exploited the ability to tune the cavity mode energy either by varying the cavity length or by changing its in-plane wave vector $k \approx (\omega/c)\sin(\theta)$, that is controlled by detection angle $\theta$ \cite{Houdre1994, Dovzhenko2018}. For example, angle-resolved reflectivity spectroscopy revealed the strong coupling with the Rabi splitting of $\sim$46 meV in MoS${_2}$ monolayer inside the Bragg microcavity \cite{Liu2015}. Similar research of both reflection and photoluminescence (PL) demonstrated a strong coupling in WSe${_2}$ and WS${_2}$ monolayers in photonic crystals \cite{Zhang2018}. Changing the cavity length in the open microcavity made it possible to observe polaritons with the Rabi splitting $\sim$20 meV in MoSe${_2}$ monolayers \cite{Dufferwiel2015} and $\sim$70 meV in WS${_2}$ monolayers \cite{Flatten2016}. 

TMDC nanotubes (NTs) were synthesized shortly after carbon NTs \cite{Iijima1991} in the early 1990s \cite{Tenne1992}. They are usually multi-walled, have a length from hundreds of nanometers to several millimeters and a characteristic diameter from tens of nanometers to several micrometers \cite{Remskar1996, Remskar2004}. We will use the same designation "NT" for both micro- and nanotubes. Tube shapes can vary from perfectly cylindrical to almost ribbon-like, depending on the internal tensile stresses \cite{Kralj2002, Enyashin2007}. In twisted tubes, in addition, the flattened cross-section rotates along the tube axis \cite{Remskar2022}. Multi-walled NTs have an indirect band structure; however, resonances of direct excitons are observed up to room temperatures  due to their giant oscillator strength and  high recombination rate \cite{Selig2016,  Shubina2019}. While individual monolayers are unstable under ambient conditions and must be protected by hBN \cite{Cadiz2017},  monolayers scrolled in NT are naturally protected by outer layers. 

Recently, many interesting phenomena have been discovered in NTs that are promising for wide application \cite{Musfeldt2020, Remskar2022}. Among them is the specific confinement of electromagnetic fields in NTs, leading to the appearance of whispering gallery modes (WGMs), the peaks of which modulate the emission spectra \cite{Kazanov2018}.
The coexistence of strong direct excitons and pronounced optical modes is very promising for the creation of polaritons in the strong coupling regime without an external cavity. However, the application of the described above detection methods with energy-momentum resolution to NTs seems to be very problematic. Providing a non-zero wave vector $k$ along the NT axis by changing the angle of incidence can hardly be combined with precise focusing on a single NT, especially in cryogenic measurements. It is also impossible to tune the optical mode energy by gradually reducing the diameter of the tube, as it is done for nanowires \cite{Sun2008}, since in conventional synthesized cylindrical NTs the cross-section is constant over a length of more than 50 $\mu$m \cite{Kazanov2020}.

To record the dispersion of polaritons, Yadgarov \textit{et al.} \cite{Yadgarov2018} used extinction measurements in ensembles of NTs of different diameters in the range of 36–110 nm, separated by centrifugation, as was previously performed for carbon NTs \cite{Graf2016}. It was noted that strong coupling can be realized if the tube diameter exceeds 80 nm \cite{Sinha2020}. With a limited number of ensembles, the ability of such an indirect method to image the dispersion of polaritons is questionable. Indisputable evidence of exciton-polaritons would be their direct observation in a single NT, which was still missing.

\begin{figure}[h]
    \centering
    \includegraphics[width=0.8\columnwidth]{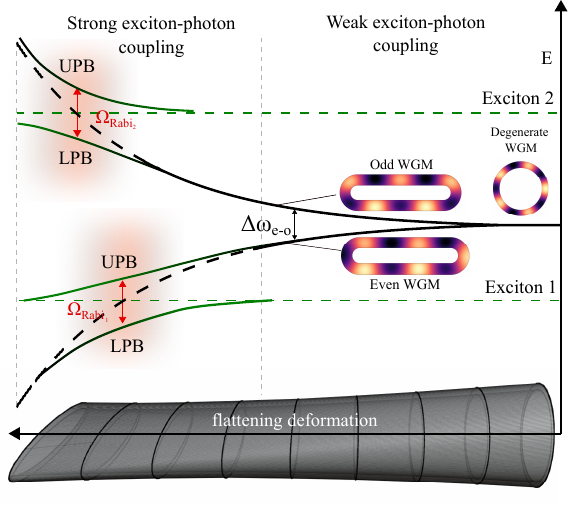}
    \caption{
    Split-mode exciton-polaritons. A doubly degenerate WGM in a cylindrical tube, as the gap between the tube walls decreases, splits into even and odd modes. When the mode reaches exciton resonance, upper and lower polariton branches are formed, exhibiting anticrossing with Rabi splitting. Black dashed lines show optical modes without interaction with excitons.
    } 
    \label{fig:dispersion}
\end{figure}

To carry out such an experiment, we propose a new concept that exploits a recently discovered phenomenon, namely, the splitting of optical modes in an NT with a flattened cross-section that features a gradual decrease in the gap between opposite walls along the NT \cite{Eliseyev2023}.
Figure~\ref{fig:dispersion} illustrates the principle of split-mode polariton formation. In a cylindrical NT the WGM is doubly degenerate and has energy of $\hbar\omega_0$. As the minor axis $b$ of the NT cross-section decreases, the gap between the walls also decreases and the evanescent tails of the electric fields begin to overlap. In analogy with the electron states in the two-well potential \cite {Landau3}, this leads to the splitting of the WGM into two non-degenerate even and odd  modes, which are symmetric and antisymmetric with respect to the major axis. Their energies read $\hbar\omega_{\rm{e,o}}=\hbar\omega_0(1 \pm t_{\rm{WGM}})$, where the odd mode has higher energy.
Note that the splitting increases exponentially as $t_{\rm{WGM}}\propto e^{-\kappa_r b}$, where $\kappa_{r}$ is the decay constant of the evanescent tails (see Supplementary for details). When an odd or even mode approaches the exciton resonance, in the case of strong coupling, an exciton-polariton is formed and Rabi splitting of the upper polariton branch (UPB) and lower polariton branch (LPB) is observed.

This work presents studies of split-mode polaritons in a single MoS$_2$ NT using micro-reflectance and micro-PL spectroscopy. It is shown that the energy of the split mode can vary over a wide range in a tube with a variable gap between the walls, which makes it possible to observe the transformation of purely optical modes into polariton branches demonstrating distinct anticrossing. On the contrary, dispersionless polaritons with stable energies are observed in NTs with a constant gap.  In both cases, we register close Rabi splitting values of $\sim 40$ meV for A-exciton-polariton and $\sim 60$ meV for B-exciton-polariton. The observed acceleration of the PL dynamics is consistent with the formation of polaritons.

\section*{Results}\label{sec2}
\subsection*{Variable wall-gap nanotube }\label{sec1}
For the experimental study, we selected a twisted MoS$_2$ NT with a total length of about one hundred microns and almost constant circumference, synthesized by a chemical vapor transfer reaction using iodine as a transport agent \cite{Remskar2022}. A common property of multi-walled NTs is internal tensile strain, which arises due to the fact that each subsequent layer must be stretched relative to the previous one in order to maintain the crystalline structure. The tensile strain promotes flattening of the cross-section and, in addition, reduces the band gap \cite{Ghorbani-Asl2013}, as occurs in atomically-thin layers under stress \cite{Peelaers2012, Conley2013, lloyd2016}.
Using Raman studies (see the Supplementary), we estimated the strain values in the NT to be $\geq$1$\%$, resulting in a red shift of A exciton towards 1.72-1.75 eV with B exciton being $\sim$140 meV higher in energy. New strain-induced positions of excitons are about 100 meV lower than in untrained flakes obtained in the same synthesis process \cite{Shubina2019}. In different regions, the NT has a different degree of strain that results in the different flattening of cross-section and rotation angle. The cross-section circumference remains constant along the NT.

\begin{figure}[t] 
    \centering
    \includegraphics[width=0.99\columnwidth]{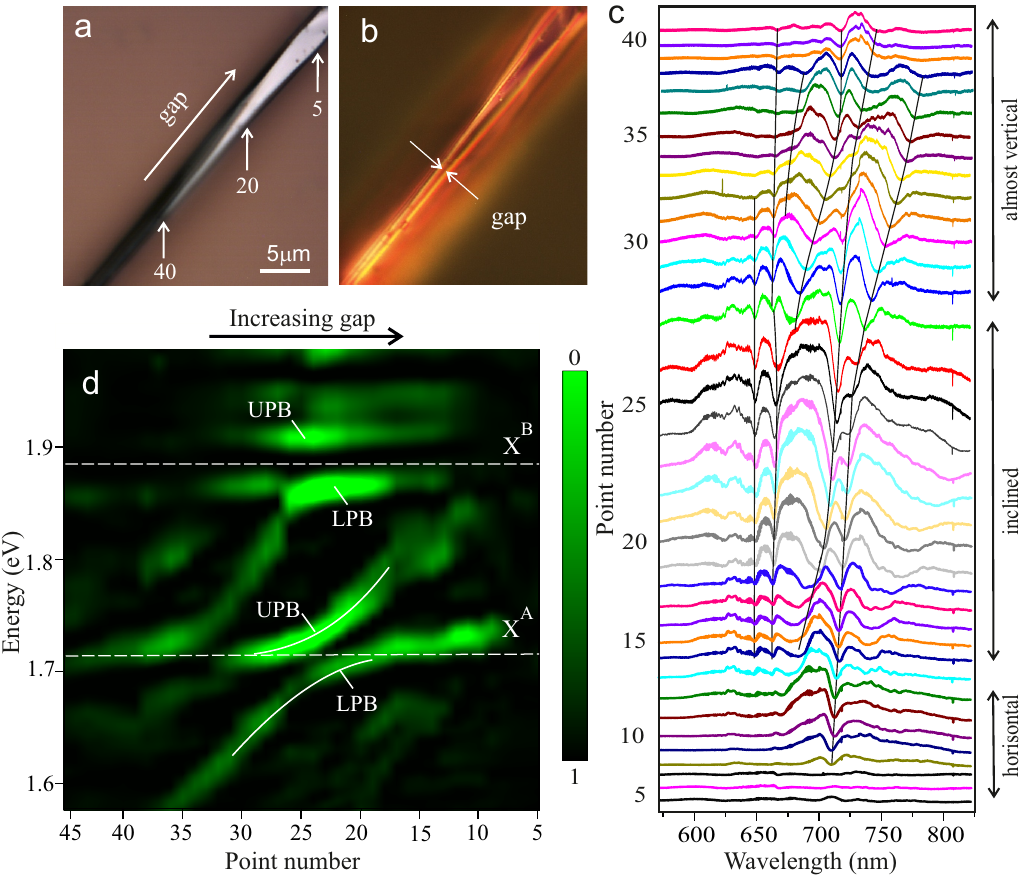}
    \caption{
    Bright-field (a) and cross-polarized dark-field (b) images of the NT with the variable gap between walls. (c) Raw reflectance spectra taken in different points along the NT (shifted vertically for clearness). (d) Map of the processed reflection spectra, where the bright green stripes represent the main dips in reflection. The dashed lines indicate an energy of excitons A and B.  The LPB and UPB are marked for both exciton-polariton. The numbering of the measurement points in (c) and (d) is the same as in (a).
    }
    \label{fig:experiment_3}
\end{figure}

Optical images of a part of the NT with a variable gap between the walls, shown in Fig. \ref{fig:experiment_3} (a,b), were obtained in different registration modes:  (a) bright-field, when light scattered perpendicularly from the surface is recorded; (b) dark-field with cross-polarization, when light scattered by inclined surfaces is recorded. The bright-field image shows the rotation of the cross-section along the NT, whereas the dark-field image shows a gradual decrease in the gap between the opposite walls of the NT. It takes place with the increase the detection point number. At the narrowest point, the gap is comparable to the thickness of the tube walls of 50-100 nm.

Figure \ref{fig:experiment_3} shows selected original (c) and processed (d) reflectance spectra measured along the NT axis as described in the Method section, with a distance between adjacent points of 1-1.5 $\mu$m. In Fig. \ref{fig:experiment_3}(c), thin lines trace main dips in the reflectance spectra. Based on the width of isolated peaks of optical modes, the quality factor of the NT cavity is estimated at $\sim$500. Far from the exciton resonances, the dips are associated with the pure optical modes excited inside the NT walls. In the points No. $30-40$, the modes (even, presumably) are observed up to the region of relative transparency (1.55 eV). The mode splitting is maximal here due to the narrowest gap. When the split modes are close to excitons in energy, which occurs in the inclined region (points No. 15-27), exciton-polaritons are formed. The reflectance in this region is approximately four times higher than in others. This is partly the result of strong exciton-photon coupling, and partly due to the dependence of the WGM mode brightness on the rotation angle of the NT cross-section \cite{Eliseyev2023}.

To get a clearer picture of the mode transformation, we removed hardware interference and background noise from the reflection spectra. In the demo display shown in Fig. \ref{fig:experiment_3} (d), green areas correspond to dips down to zero in the original reflectance spectra. A clear anticrossing of LPB and UPB is observed when one of the split modes approaches exciton A. The other mode creates anticrossing branches at exciton B.
The UPB at exciton B is less pronounced due to high absorption in the spectral region above exciton B.

In the region where the gap between the walls increases (points No. 5-15) and respectively the strain value becomes smaller, a blue shift of the A exciton resonance is observed, which affects the dispersion of polariton branches, bending them towards higher energy compared to the average.
The experimentally recorded Rabi splittings between the LPB and UPB are found to be about 40 meV for exciton A (1.74 eV) and 60 meV for exciton B (1.88 eV), which has higher oscillator strength. These values indicate a strong coupling regime and provide a sufficient ratio of $\gtrsim 3$ to the width of the reflection dip to observe the full picture of polaritons using reflectance spectroscopy.

To analyze the evolution of optical modes leading to the formation of polariton, we consider a simplified theoretical model.
In the weakly deformed NT, the electric field of WGMs polarized along the NT axis is described by $E^{\rm (e)}_m \propto {\rm cos}(m\varphi)$ and $E^{\rm (o)}_m \propto {\rm sin}(m\varphi)$, where $e$ and $o$ denote even and odd modes, $m$ is the angular number, and the angle $\phi$ is measured from the major axis of the flattened cross-section. If light is incident normally on the NT axis, the back-scattering intensity near the resonance of the even mode $\omega^{(m)}_{\rm e}$ has the form
\begin{equation}
R(\omega,\alpha) \approx \bigg |\frac{A^{(m)} {\rm cos}^2(m[\alpha+\frac{\pi}{2}])}{\omega-\omega^{(m)}_{\rm e} +{\rm i}\Gamma^{(m)}} + C^{(m)} \bigg |^2,
\end{equation}\label{eq:Reflection_full}
where $\alpha$ is the angle between the minor axis of cross-section and light propagation direction, $\Gamma^{(m)}$ is the decay rate of the mode, which comprises both the radiative and non-radiative contributions, $A^{(m)}$ characterizes the strength of the resonance, and $ C^{(m)}$ describes the contribution of all other modes. For an odd WGM, in Eq.~(1) $\cos$ should be replaced by $\sin$.  The trigonometric function in the numerator originates from the convolution of the incident field with the field distribution of the mode. This leads to a change in the visibility of modes in the spectra when the cross section rotates. (The details on the derivation are given in the Supplementary.) 

We performed numerical calculations to elucidate the evolution of modes taking into account exciton resonances in the strong coupling regime using Comsol Multiphysics as described in the Method section.
The  dielectric response of nanotube wall was taken in Drude-Lorentz form with two resonances of A and B excitons:
\begin{equation}
\varepsilon = \varepsilon_b\bigg[1 + \sum_{i={\rm A,B}} \frac{ \Omega^2_i}{(\omega^2_{i}-\omega^2-{\rm i}\omega\Gamma_i)} \bigg].
\end{equation}
Here, $i$ denotes excitons A and B, $\Omega_{i}$ is the Rabi splitting in the bulk material, $\Gamma_i$ is the non-radiative broadening of exciton resonances, $\varepsilon_b$ ($\sim 12.3$) is the background dielectric constant of MoS$_2$. We assume that the NT cross-section has a racetrack shape with a constant circumference ($\sim$5.5 $\mu$m). The NT wall thickness (50 nm) and rotation angle ($65^{\circ}$) are also fixed in the calculations.

\begin{figure}[t]
    \centering
    \includegraphics[width=0.99\columnwidth]{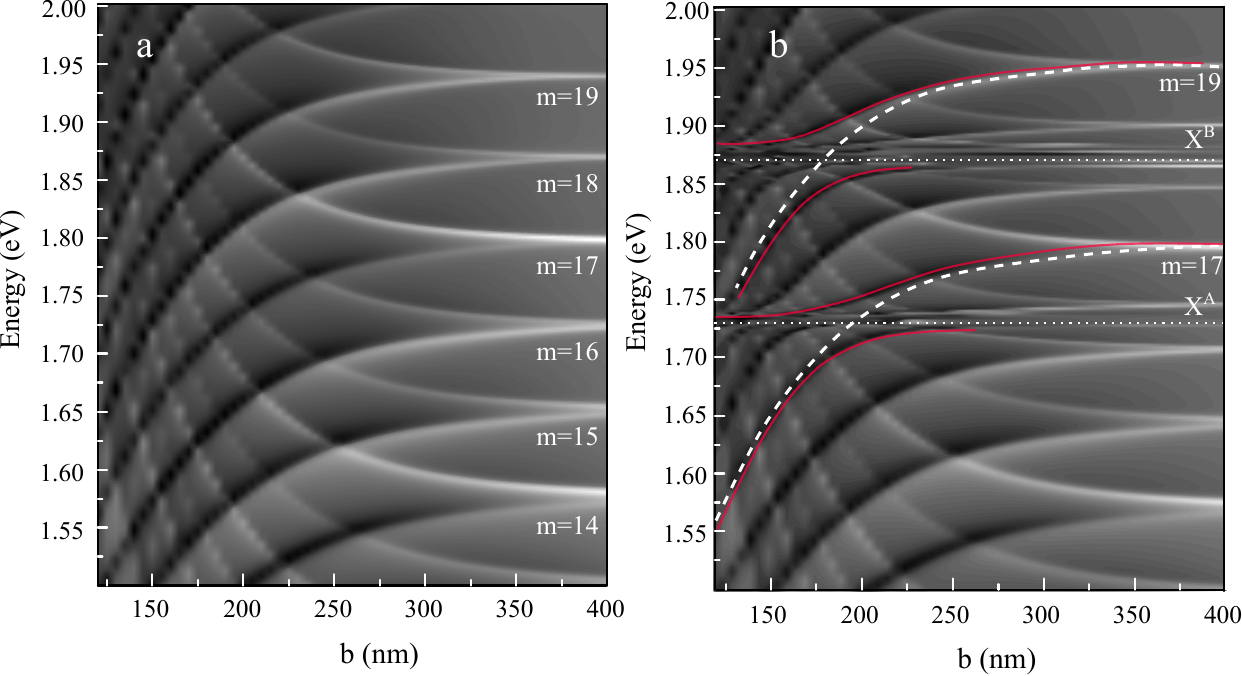}
\caption
{Splitting and evolution of even and odd optical WGMs in reflection with NT flattening described by the minor axis $b$ of cross-section. (a) Bare WGMs without interaction with the resonances of A and B excitons. A change in the color of a curve from light to dark shows the depth of the reflection dip. (b) WGMs and exciton resonances, depicted by white dotted lines, in the strong coupling regime generating polaritons. The red lines indicate the LPB and UPB with Rabi splitting, which arise from the even modes near excitons A and B. The dashed white lines show the same modes without interaction.}
    \label{fig:ModesEvolution}
\end{figure}

Figure \ref{fig:ModesEvolution}(a) shows the behavior of WGMs without interaction with exciton resonances.  As the cross-section flattens, all even modes, curved downward, decrease their energy, and all odd modes, curved upward, increase energy. When the minor axis $b$ of the cross-section becomes less than half the wavelength of the exciting light, the odd-mode resonances begin to quench. At a double-well potential \cite{Landau3}, these antisymmetric modes have electric field in antiphase on opposite sides of the cross-section. Therefore, their coupling to the free waves is decreased as the gap between the walls narrows. In contrast, for even modes, which have electric field of the same phase on the opposite sides, the coupling to the free waves increases. This explains why only even, downward-curved modes are retained in the experimental spectra of highly flattened NTs.

The case of strong coupling between the optical modes and the excitons is shown in Fig. \ref{fig:ModesEvolution}(b).
In the absence of the cross-section deformation, WGMs shift their energies slightly near the exciton resonances due to the changes in the dielectric background. As the axis $b$ decreases to 200-250 nm, stronger even modes reach exciton resonances from above and form exciton-polaritons. For clarity, we have highlighted with red lines the upper and lower polariton branches originating from the WGMs with angular numbers $m=17$ and $m=19$. 
The anticrossing with $\Omega_{\rm Rabi}$ of $\sim35$ meV and $\sim 55$ meV for excitons A and B is quite close to the experimental results. The general picture of the calculated modes and their slopes also satisfactorily reproduce the experimental spectra.

\subsection*{Constant wall-gap nanotube}\label{sec1}

\begin{figure}[b]
    \centering
\includegraphics[width=0.99\columnwidth]{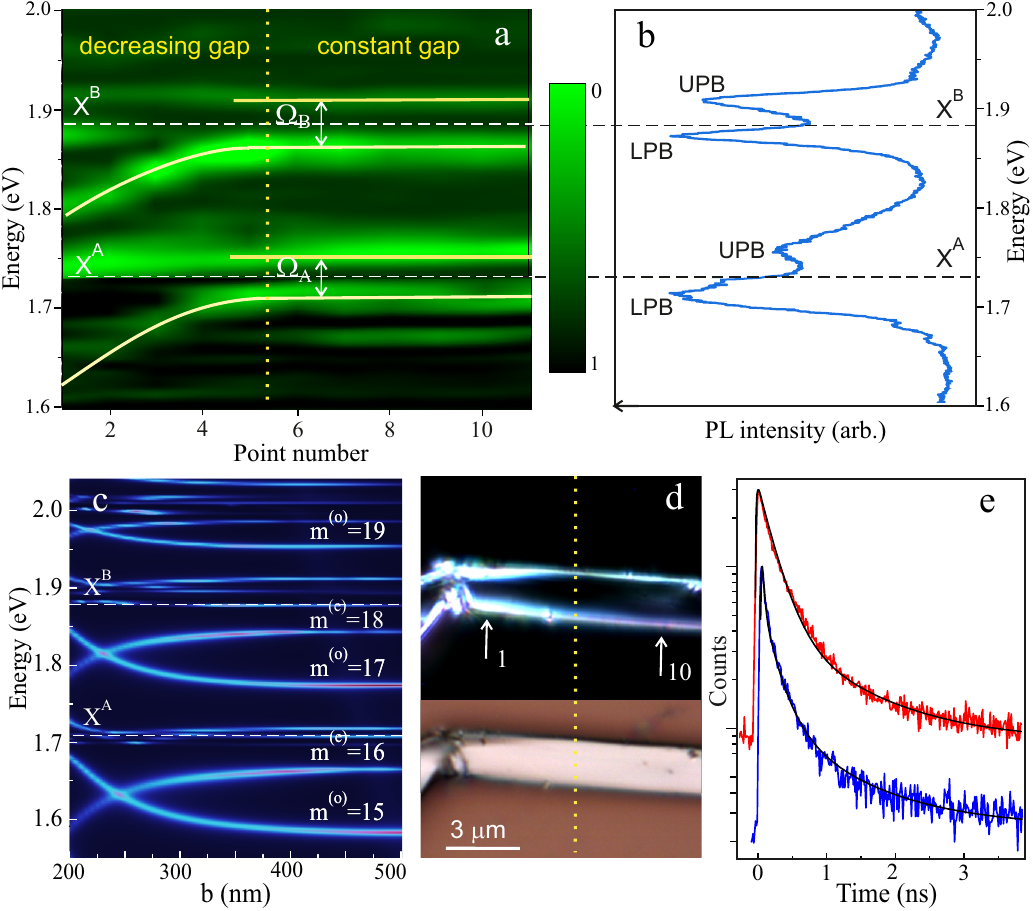}
\caption
{(a) Map of processed reflectance spectra measured in an NT with a constant gap. Green lines correspond to the main reflection dips, white dotted lines correspond to the energies of A and B excitons, and yellow lines highlight the polariton modes. 
(b) The PL spectrum measured in the constant-gap region, showing four peaks associated with stable polariton branches. (c) Simulation of modes at the rotation angle of $10^{\circ}$ as a function of the minor axis $b$; $m^{(e,o)}$ denotes the angular number of even or odd modes. (d) Dark-field (top) and bright-field (bottom) optical images of this NT. (e) PL decay curves measured in  A exciton LPB peak 
in (b) (blue) and a cylindrical tube (red). Vertical yellow lines  separate the constant and decreasing gap regions.
}
    \label{fig:Figure4}
\end{figure}

Now let us focus on another NT, which has an extended flat region with a small rotation angle and a constant gap between the walls (see Fig. \ref{fig:Figure4}). Map the reflection spectra presented in Fig. \ref{fig:Figure4} (a) shows optical modes of constant energies along the flat region. They are grouped into pairs with mode spacing in each pair very close to the Rabi splitting for A and B exciton-polaritons. PL spectra measured in this region exhibit four peaks that perfectly correspond to the modes in reflection (Fig. \ref{fig:Figure4}c). 
This picture corresponds to the dispersionless polaritons with constant energy, which were theoretically considered for TMDC monolayers  \cite{Alpeggiani2018}. 
The parameters of these polaritons are stable along the large segment of the NT, where its geometry does not change. 

Calculations of optical modes for this case were carried out using almost the same model as described above, but assuming a close to zero cross-section rotation angle. We did not divide the NT into two regions, but gradually changed the value of $b$ from 500 nm to 200 nm (Fig. \ref{fig:Figure4}). For the same circumference of $\sim$5.5 $\mu$m, the found optical modes have the angular numbers $m$ as in Fig. \ref{fig:ModesEvolution}a. However, there is a remarkable difference due to the small rotation angle. At degrees less than $10^{\circ}$, only one of the two modes (even or odd) is excited, and this occurs in antiphase to each adjacent WGM pair. 

Comparison of experimental and calculated pictures in Fig. \ref{fig:Figure4} (a, c) shows that the observed stability of mode energies and their separation can be realized only if the axis $b$ exceeds 350 nm. Although the change in mode energies above this value is weak, it can certainly be resolved experimentally. Thus, the gap value is most likely constant and equal $\sim$400 nm. In another part of the NT, close to its kink, when the deformation increases and the gap is narrowed, the even modes begin to shift towards lower energies in both experimental and calculated spectra, as it occurs in the NT with the variable gap (Fig. \ref{fig:experiment_3}). In the case of completely flattened NT with the shape of a ribbon, it is possible to form the polaritons on Fabry-P$\acute{e}$rot modes as it has been observed in thin WS$_2$ flakes \cite{Wang2016}. 

In addition, we performed time-resolved photoluminescence (TRPL) measurements as described in the Method section. Figure \ref{fig:Figure4} (e) shows a PL decay curve for the peak of A exciton LPB in Fig. \ref{fig:Figure4} (b) together with a curve measured at the same energy in a reference cylindrical NT that exhibits conventional WGMs. Both curves contain fast components with characteristic decay times of less than 20 ps and slow components with decay times of hundreds of ps. In the NT under study, the contribution of the fast component to the integral PL is approximately 30 times higher than in the reference NT.  Since the fast component reflects the dynamics of polaritons \cite{Lundt2017}, such a difference indirectly confirms the strong coupling regime in the studied NTs.

\subsection*{In conclusion}\label{sec3}
The discovery of anticrossing of polariton branches is usually considered as indisputable evidence of the existence of polaritons in the strong coupling regime. We were able to obtain this result using the change in the energy of split modes in NT with a variable gap between the walls in a flattened cross-section. This approach represents an optical-mechanical method for controlling the interaction of exciton resonances with a set of optical modes excited in NT.

Another important discovery was the observation of polaritons with stable energy in an NT with a constant gap in an unchanged cross-section. The general characteristics of such  dispersionless polaritons are in a good agreement with those found for exciton-polaritons in the NT with the variable gap, which excludes another explanation. Note that polaritons of this type were realized at a relatively large gap between NT walls ($\sim$400 nm), which indicates the possibility of their excitation in almost cylindrical MoS$_2$ NTs.

In both cases, the experimentally observed pattern of optical modes and their transformation into polariton branches are consistent with the proposed theoretical description. The simulation performed reproduces polaritons with Rabi splitting values close to experimental ones and makes it possible to reconstruct the geometric parameters of the NTs under study. Taken together, our experimental and theoretical results confirm the possibility of producing polaritons in the strong coupling regime in individual TMDC NTs, making them promising candidates for nanophotonics.

\section*{Methods}\label{Methods}
\subsection*{Optical studies}
Optical measurements of MoS$_2$ NTs were done at 8 K. The Si/SiO$_2$ substrate with the NTs on its surface was mounted in the ST-500-Attocube (Janis) cryostat, containing the three-coordinate piezo-driver with an accuracy in positioning of $\sim$20 nm, located in the cold zone of the cryostat. This provides high mechanical stability and vibration isolation necessary for studying such tiny objects. For PL excitation we used a semiconductor laser PILAS 405 nm (Advanced Laser Systems) with a 40 ps pulse duration and 80 MHz repetition rate. Focusing the laser light using an 100x apochromatic objective (Mitutoyo plan apochromat, NA = 0.7) provided spot size less than 1 $\mu$m, which corresponds to $\sim$ 6 W/cm$^2$ (10 nW/$\mu$m$^2$)nm power density on the sample surface. The PL signal was collected by the same objective in a confocal setup and directed onto the entrance slit of an SP-2500 spectrometer with cooled PyLoN CCD detector (Princeton Instruments). To measure the time resolved histogram of PL a single-photon avalanche photodiode PDM 100 (Micro Photon Devices) with time resolution $\sim$ 40 ps was used. For blocking the scattered laser radiation, a band pass interference filter was used. 

The reflection spectra along the NTs were measured on the same optical setup using a halogen lamp coupled to a single-mode optical fiber with subsequent collimation. In this mode, it was possible to achieve high spatial resolution when scanning the NTs. To highlight features, the reflection spectra were subjected to additional processing. First, the reflection spectrum was smoothed using a low-pass Fourier filter to get rid of unwanted noise. Secondly, the moving-average background was subtracted from the reflection spectrum, which made it possible to enhance the visibility of exciton and polariton reflection dips against of other effects.

Raman spectra were obtained in backscattering geometry at room temperature. For these investigations, we used a Horiba LabRAM HREvo UV-VIS-NIR-Open spectrometer (Horiba, Lille, France) with confocal optics. Olympus MPLN100x (Olympus, Tokyo, Japan) objective lens (NA = 0.9) focused the laser beam into a spot $\sim$1 $\mu$m in diameter. As an excitation source, a Nd:YAG laser (Torus Laser Quantum, Stockport, UK) with $\lambda = 532$ nm  was used. To avoid damage and heating of the NTs, the laser power on its surface was limited to 200 $\mu$W.

Optical images of the NTs in different polarizations were obtained using a Nikon ECLIPSE LV150 (Nikon Europe, Amsterdam, Netherlands) industrial microscope operating in bright-field and dark-field modes. The sample under study was illuminated using an LED lamp passed through a linear polarizer while long-exposure detection was carried out in orthogonal polarization. This method made it possible to see features of NTs that were inaccessible with the conventional method.

\subsection*{Comsol Multiphysics calculations} 
To simulate the reflection and PL map, we divided computational domain into two parts, outside part with conventional refractive index of air, and inside NT domain with refractive index using Drude-Lorentz dispersion model with A and B exciton resonances. The exciton line broadening was chosen to be 5 meV to resolve the rich pattern of optical modes of in the TMDC NT. We assumed a wave vector along the NT axis $k_z=0$, because $k_z\neq 0$ shifts the energies of optical modes in the NT much less than the flattening effects. To neglect reflection from computational domain boundaries, we added a boundary box with perfectly matched layer. To excite optical modes, we put a plane-wave source incident on the 2D cross-section of NT with polarization along the NT axis. To find the back scattered signal, we summed the reflected electric field far away from the NT along a line, while to find the PL signal we calculated the integral of the squared electromagnetic field inside the NT walls.

\subsection*{Conflict of Interest}
The authors declare no conflict of interest.

\subsection*{Author contribution}
All authors contributed to this article. A.I.G., M.V.R., I.A.E. and A.A.T. carried out optical measurements and analyzed data; D.R.K. and A.V.P. performed theoretical calculations; M.R. carried out the growth of nanotubes; T.V.S. proposed the physical model, wrote the final version, and supervised this research. All authors gave approval for the final version of the manuscript. 

\subsection*{Acknowledgements}
This work was supported in part by the Russian Science Foundation project no. 23-12-00300 (A.I.G. - reflection study, T.V.S. - physical model). We thank B. Borodin for transferring and positioning the nanotubes on the substrate and A. Veretennikov for his assistance in data processing.

\section*{Data availability}
The data supporting the findings of this study are available upon reasonable request.

\backmatter

\bmhead{Supplementary information}

Accompanying supplementary file is added.



\begin{thebibliography}{44}
\ifx \bisbn   \undefined \def \bisbn  #1{ISBN #1}\fi
\ifx \binits  \undefined \def \binits#1{#1}\fi
\ifx \bauthor  \undefined \def \bauthor#1{#1}\fi
\ifx \batitle  \undefined \def \batitle#1{#1}\fi
\ifx \bjtitle  \undefined \def \bjtitle#1{#1}\fi
\ifx \bvolume  \undefined \def \bvolume#1{\textbf{#1}}\fi
\ifx \byear  \undefined \def \byear#1{#1}\fi
\ifx \bissue  \undefined \def \bissue#1{#1}\fi
\ifx \bfpage  \undefined \def \bfpage#1{#1}\fi
\ifx \blpage  \undefined \def \blpage #1{#1}\fi
\ifx \burl  \undefined \def \burl#1{\textsf{#1}}\fi
\ifx \doiurl  \undefined \def \doiurl#1{\url{https://doi.org/#1}}\fi
\ifx \betal  \undefined \def \betal{\textit{et al.}}\fi
\ifx \binstitute  \undefined \def \binstitute#1{#1}\fi
\ifx \binstitutionaled  \undefined \def \binstitutionaled#1{#1}\fi
\ifx \bctitle  \undefined \def \bctitle#1{#1}\fi
\ifx \beditor  \undefined \def \beditor#1{#1}\fi
\ifx \bpublisher  \undefined \def \bpublisher#1{#1}\fi
\ifx \bbtitle  \undefined \def \bbtitle#1{#1}\fi
\ifx \bedition  \undefined \def \bedition#1{#1}\fi
\ifx \bseriesno  \undefined \def \bseriesno#1{#1}\fi
\ifx \blocation  \undefined \def \blocation#1{#1}\fi
\ifx \bsertitle  \undefined \def \bsertitle#1{#1}\fi
\ifx \bsnm \undefined \def \bsnm#1{#1}\fi
\ifx \bsuffix \undefined \def \bsuffix#1{#1}\fi
\ifx \bparticle \undefined \def \bparticle#1{#1}\fi
\ifx \barticle \undefined \def \barticle#1{#1}\fi
\bibcommenthead
\ifx \bconfdate \undefined \def \bconfdate #1{#1}\fi
\ifx \botherref \undefined \def \botherref #1{#1}\fi
\ifx \url \undefined \def \url#1{\textsf{#1}}\fi
\ifx \bchapter \undefined \def \bchapter#1{#1}\fi
\ifx \bbook \undefined \def \bbook#1{#1}\fi
\ifx \bcomment \undefined \def \bcomment#1{#1}\fi
\ifx \oauthor \undefined \def \oauthor#1{#1}\fi
\ifx \citeauthoryear \undefined \def \citeauthoryear#1{#1}\fi
\ifx \endbibitem  \undefined \def \endbibitem {}\fi
\ifx \bconflocation  \undefined \def \bconflocation#1{#1}\fi
\ifx \arxivurl  \undefined \def \arxivurl#1{\textsf{#1}}\fi
\csname PreBibitemsHook\endcsname

\bibitem[\protect\citeauthoryear{Evans and Young}{1965}]{Evans1965}
\begin{botherref}
\oauthor{\bsnm{Evans}, \binits{B.L.}},
\oauthor{\bsnm{Young}, \binits{P.A.}}:
Optical absorption and dispersion in molybdenum disulphide.
Proc. R. Soc. Lond. A
\textbf{285}
(1965)
\end{botherref}
\endbibitem

\bibitem[\protect\citeauthoryear{Wang et~al.}{2018}]{Wang2018}
\begin{barticle}
\bauthor{\bsnm{Wang}, \binits{G.}},
\bauthor{\bsnm{Chernikov}, \binits{A.}},
\bauthor{\bsnm{Glazov}, \binits{M.M.}},
\bauthor{\bsnm{Heinz}, \binits{T.F.}},
\bauthor{\bsnm{Marie}, \binits{X.}},
\bauthor{\bsnm{Amand}, \binits{T.}},
\bauthor{\bsnm{Urbaszek}, \binits{B.}}:
\batitle{Colloquium: Excitons in atomically thin transition metal dichalcogenides}.
\bjtitle{Rev. Mod. Phys}
\bvolume{90}(\bissue{2}),
\bfpage{021001}
(\byear{2018})
\end{barticle}
\endbibitem

\bibitem[\protect\citeauthoryear{Imamoglu et~al.}{1996}]{Imamoglu1996}
\begin{barticle}
\bauthor{\bsnm{Imamoglu}, \binits{A.}},
\bauthor{\bsnm{Ram}, \binits{R.J.}},
\bauthor{\bsnm{Pau}, \binits{S.}},
\bauthor{\bsnm{Yamamoto}, \binits{Y.}}:
\batitle{Nonequilibrium condensates and lasers without inversion: Exciton-polariton lasers}.
\bjtitle{Phys. Rev. A}
\bvolume{53},
\bfpage{4250}--\blpage{4253}
(\byear{1996})
\end{barticle}
\endbibitem

\bibitem[\protect\citeauthoryear{Ghosh and Liew}{2020}]{Ghosh2020}
\begin{barticle}
\bauthor{\bsnm{Ghosh}, \binits{S.}},
\bauthor{\bsnm{Liew}, \binits{T.C.H.}}:
\batitle{Quantum computing with exciton-polariton condensates}.
\bjtitle{npj Quantum Information}
\bvolume{6}(\bissue{1}),
\bfpage{16}
(\byear{2020})
\end{barticle}
\endbibitem

\bibitem[\protect\citeauthoryear{Kim and Yamamoto}{2017}]{Kim2017}
\begin{bbook}
\bauthor{\bsnm{Kim}, \binits{N.Y.}},
\bauthor{\bsnm{Yamamoto}, \binits{Y.}}:
\bbtitle{Exciton-Polariton Quantum Simulators},
pp. \bfpage{91}--\blpage{121}.
\bpublisher{Springer},
\blocation{Cham}
(\byear{2017})
\end{bbook}
\endbibitem

\bibitem[\protect\citeauthoryear{Kavokin et~al.}{2008}]{Kavokin2008}
\begin{bbook}
\bauthor{\bsnm{Kavokin}, \binits{A.V.}},
\bauthor{\bsnm{Baumberg}, \binits{J.J.}},
\bauthor{\bsnm{Malpeuch}, \binits{G.}},
\bauthor{\bsnm{Laussy}, \binits{F.P.}}:
\bbtitle{Microcavities}.
\bpublisher{Oxford University Press},
\blocation{UK}
(\byear{2008})
\end{bbook}
\endbibitem

\bibitem[\protect\citeauthoryear{Luo et~al.}{2023}]{Luo2023}
\begin{botherref}
\oauthor{\bsnm{Luo}, \binits{S.}},
\oauthor{\bsnm{Zhou}, \binits{H.}},
\oauthor{\bsnm{Zhang}, \binits{L.}},
\oauthor{\bsnm{Chen}, \binits{Z.}}:
{Nanophotonics of microcavity exciton–polaritons}.
Applied Physics Reviews
\textbf{10}(1)
(2023).
011316
\end{botherref}
\endbibitem

\bibitem[\protect\citeauthoryear{Wei et~al.}{2023}]{Wei2023}
\begin{barticle}
\bauthor{\bsnm{Wei}, \binits{K.}},
\bauthor{\bsnm{Liu}, \binits{Q.}},
\bauthor{\bsnm{Tang}, \binits{Y.}},
\bauthor{\bsnm{Ye}, \binits{Y.}},
\bauthor{\bsnm{Xu}, \binits{Z.}},
\bauthor{\bsnm{Jiang}, \binits{T.}}:
\batitle{Charged biexciton polaritons sustaining strong nonlinearity in {2D} semiconductor-based nanocavities}.
\bjtitle{Nature Communications}
\bvolume{14}(\bissue{1}),
\bfpage{5310}
(\byear{2023})
\end{barticle}
\endbibitem

\bibitem[\protect\citeauthoryear{Weisbuch et~al.}{1992}]{Weisbuch1992}
\begin{barticle}
\bauthor{\bsnm{Weisbuch}, \binits{C.}},
\bauthor{\bsnm{Nishioka}, \binits{M.}},
\bauthor{\bsnm{Ishikawa}, \binits{A.}},
\bauthor{\bsnm{Arakawa}, \binits{Y.}}:
\batitle{Observation of the coupled exciton-photon mode splitting in a semiconductor quantum microcavity}.
\bjtitle{Phys. Rev. Lett.}
\bvolume{69},
\bfpage{3314}--\blpage{3317}
(\byear{1992})
\end{barticle}
\endbibitem

\bibitem[\protect\citeauthoryear{Kulakovskii et~al.}{2000}]{Kulakovskii2000}
\begin{barticle}
\bauthor{\bsnm{Kulakovskii}, \binits{V.D.}},
\bauthor{\bsnm{Tartakovskii}, \binits{A.I.}},
\bauthor{\bsnm{Krizhanovskii}, \binits{D.N.}},
\bauthor{\bsnm{Armitage}, \binits{A.}},
\bauthor{\bsnm{Roberts}, \binits{J.S.}},
\bauthor{\bsnm{Skolnick}, \binits{M.S.}}:
\batitle{Two-dimensional excitonic polaritons and their interaction}.
\bjtitle{Physics-Uspekhi}
\bvolume{43}(\bissue{8}),
\bfpage{853}
(\byear{2000})
\end{barticle}
\endbibitem

\bibitem[\protect\citeauthoryear{Liu et~al.}{2015}]{Liu2015}
\begin{barticle}
\bauthor{\bsnm{Liu}, \binits{X.}},
\bauthor{\bsnm{Galfsky}, \binits{T.}},
\bauthor{\bsnm{Sun}, \binits{Z.}},
\bauthor{\bsnm{Xia}, \binits{F.}},
\bauthor{\bsnm{Lin}, \binits{E.-c.}},
\bauthor{\bsnm{Lee}, \binits{Y.-H.}},
\bauthor{\bsnm{K{\'e}na-Cohen}, \binits{S.}},
\bauthor{\bsnm{Menon}, \binits{V.M.}}:
\batitle{Strong light--matter coupling in two-dimensional atomic crystals}.
\bjtitle{Nature Photonics}
\bvolume{9}(\bissue{1}),
\bfpage{30}--\blpage{34}
(\byear{2015})
\end{barticle}
\endbibitem

\bibitem[\protect\citeauthoryear{Dufferwiel et~al.}{2015}]{Dufferwiel2015}
\begin{barticle}
\bauthor{\bsnm{Dufferwiel}, \binits{S.}},
\bauthor{\bsnm{Schwarz}, \binits{S.}},
\bauthor{\bsnm{Withers}, \binits{F.}},
\bauthor{\bsnm{Trichet}, \binits{A.A.P.}},
\bauthor{\bsnm{Li}, \binits{F.}},
\bauthor{\bsnm{Sich}, \binits{M.}},
\bauthor{\bsnm{Del~Pozo-Zamudio}, \binits{O.}},
\bauthor{\bsnm{Clark}, \binits{C.}},
\bauthor{\bsnm{Nalitov}, \binits{A.}},
\bauthor{\bsnm{Solnyshkov}, \binits{D.D.}},
\bauthor{\bsnm{Malpuech}, \binits{G.}},
\bauthor{\bsnm{Novoselov}, \binits{K.S.}},
\bauthor{\bsnm{Smith}, \binits{J.M.}},
\bauthor{\bsnm{Skolnick}, \binits{M.S.}},
\bauthor{\bsnm{Krizhanovskii}, \binits{D.N.}},
\bauthor{\bsnm{Tartakovskii}, \binits{A.I.}}:
\batitle{Exciton--polaritons in van der {W}aals heterostructures embedded in tunable microcavities}.
\bjtitle{Nature Communications}
\bvolume{6}(\bissue{1}),
\bfpage{8579}
(\byear{2015})
\end{barticle}
\endbibitem

\bibitem[\protect\citeauthoryear{Flatten et~al.}{2016}]{Flatten2016}
\begin{barticle}
\bauthor{\bsnm{Flatten}, \binits{L.C.}},
\bauthor{\bsnm{He}, \binits{Z.}},
\bauthor{\bsnm{Coles}, \binits{D.M.}},
\bauthor{\bsnm{Trichet}, \binits{A.A.P.}},
\bauthor{\bsnm{Powell}, \binits{A.W.}},
\bauthor{\bsnm{Taylor}, \binits{R.A.}},
\bauthor{\bsnm{Warner}, \binits{J.H.}},
\bauthor{\bsnm{Smith}, \binits{J.M.}}:
\batitle{Room-temperature exciton-polaritons with two-dimensional {W}{S}$_{2}$}.
\bjtitle{Scientific Reports}
\bvolume{6}(\bissue{1}),
\bfpage{33134}
(\byear{2016})
\end{barticle}
\endbibitem

\bibitem[\protect\citeauthoryear{Zhang et~al.}{2018}]{Zhang2018}
\begin{barticle}
\bauthor{\bsnm{Zhang}, \binits{L.}},
\bauthor{\bsnm{Gogna}, \binits{R.}},
\bauthor{\bsnm{Burg}, \binits{W.}},
\bauthor{\bsnm{Tutuc}, \binits{E.}},
\bauthor{\bsnm{Deng}, \binits{H.}}:
\batitle{Photonic-crystal exciton-polaritons in monolayer semiconductors}.
\bjtitle{Nature Communications}
\bvolume{9}(\bissue{1}),
\bfpage{713}
(\byear{2018})
\end{barticle}
\endbibitem

\bibitem[\protect\citeauthoryear{Schneider et~al.}{2018}]{Schneider2018}
\begin{barticle}
\bauthor{\bsnm{Schneider}, \binits{C.}},
\bauthor{\bsnm{Glazov}, \binits{M.M.}},
\bauthor{\bsnm{Korn}, \binits{T.}},
\bauthor{\bsnm{H{\"o}fling}, \binits{S.}},
\bauthor{\bsnm{Urbaszek}, \binits{B.}}:
\batitle{Two-dimensional semiconductors in the regime of strong light-matter coupling}.
\bjtitle{Nature Communications}
\bvolume{9}(\bissue{1}),
\bfpage{2695}
(\byear{2018})
\end{barticle}
\endbibitem

\bibitem[\protect\citeauthoryear{Lackner et~al.}{2021}]{Lackner2021}
\begin{barticle}
\bauthor{\bsnm{Lackner}, \binits{L.}},
\bauthor{\bsnm{Dusel}, \binits{M.}},
\bauthor{\bsnm{Egorov}, \binits{O.A.}},
\bauthor{\bsnm{Han}, \binits{B.}},
\bauthor{\bsnm{Knopf}, \binits{H.}},
\bauthor{\bsnm{Filenberger}, \binits{F.}},
\bauthor{\bsnm{Schr{\"o}der}, \binits{S.}},
\bauthor{\bsnm{Watanabe}, \binits{K.}},
\bauthor{\bsnm{Taniguchi}, \binits{T.}},
\bauthor{\bsnm{Tongay}, \binits{S.}},
\bauthor{\bsnm{Anton-Solanas}, \binits{C.}},
\bauthor{\bsnm{H{\"o}fling}, \binits{S.}},
\bauthor{\bsnm{Schneider}, \binits{C.}}:
\batitle{Tunable exciton-polaritons emerging from {W}{S}$_{2}$ monolayer excitons in a photonic lattice at room temperature}.
\bjtitle{Nature Communications}
\bvolume{12}(\bissue{1}),
\bfpage{4933}
(\byear{2021})
\end{barticle}
\endbibitem

\bibitem[\protect\citeauthoryear{Houdr\'e et~al.}{1994}]{Houdre1994}
\begin{barticle}
\bauthor{\bsnm{Houdr\'e}, \binits{R.}},
\bauthor{\bsnm{Weisbuch}, \binits{C.}},
\bauthor{\bsnm{Stanley}, \binits{R.P.}},
\bauthor{\bsnm{Oesterle}, \binits{U.}},
\bauthor{\bsnm{Pellandini}, \binits{P.}},
\bauthor{\bsnm{Ilegems}, \binits{M.}}:
\batitle{Measurement of cavity-polariton dispersion curve from angle-resolved photoluminescence experiments}.
\bjtitle{Phys. Rev. Lett.}
\bvolume{73},
\bfpage{2043}--\blpage{2046}
(\byear{1994})
\end{barticle}
\endbibitem

\bibitem[\protect\citeauthoryear{Dovzhenko et~al.}{2018}]{Dovzhenko2018}
\begin{barticle}
\bauthor{\bsnm{Dovzhenko}, \binits{D.S.}},
\bauthor{\bsnm{Ryabchuk}, \binits{S.V.}},
\bauthor{\bsnm{Rakovich}, \binits{Y.P.}},
\bauthor{\bsnm{Nabiev}, \binits{I.R.}}:
\batitle{Light–matter interaction in the strong coupling regime: configurations{,} conditions{,} and applications}.
\bjtitle{Nanoscale}
\bvolume{10},
\bfpage{3589}--\blpage{3605}
(\byear{2018})
\end{barticle}
\endbibitem

\bibitem[\protect\citeauthoryear{Iijima}{1991}]{Iijima1991}
\begin{barticle}
\bauthor{\bsnm{Iijima}, \binits{S.}}:
\batitle{Helical microtubules of graphitic carbon}.
\bjtitle{Nature}
\bvolume{354}(\bissue{6348}),
\bfpage{56}--\blpage{58}
(\byear{1991})
\end{barticle}
\endbibitem

\bibitem[\protect\citeauthoryear{Tenne et~al.}{1992}]{Tenne1992}
\begin{barticle}
\bauthor{\bsnm{Tenne}, \binits{R.}},
\bauthor{\bsnm{Margulis}, \binits{L.}},
\bauthor{\bsnm{Genut}, \binits{M.}},
\bauthor{\bsnm{Hodes}, \binits{G.}}:
\batitle{Polyhedral and cylindrical structures of tungsten disulphide}.
\bjtitle{Nature}
\bvolume{360}(\bissue{6403}),
\bfpage{444}--\blpage{446}
(\byear{1992})
\end{barticle}
\endbibitem

\bibitem[\protect\citeauthoryear{Remskar et~al.}{1996}]{Remskar1996}
\begin{barticle}
\bauthor{\bsnm{Remskar}, \binits{M.}},
\bauthor{\bsnm{Skraba}, \binits{Z.}},
\bauthor{\bsnm{Cléton}, \binits{F.}},
\bauthor{\bsnm{Sanjinés}, \binits{R.}},
\bauthor{\bsnm{Lévy}, \binits{F.}}:
\batitle{{MoS$_2$ as microtubes}}.
\bjtitle{Applied Physics Letters}
\bvolume{69}(\bissue{3}),
\bfpage{351}--\blpage{353}
(\byear{1996})
\end{barticle}
\endbibitem

\bibitem[\protect\citeauthoryear{Remskar}{2004}]{Remskar2004}
\begin{barticle}
\bauthor{\bsnm{Remskar}, \binits{M.}}:
\batitle{Inorganic nanotubes}.
\bjtitle{Advanced Materials}
\bvolume{16}(\bissue{17}),
\bfpage{1497}--\blpage{1504}
(\byear{2004})
\end{barticle}
\endbibitem

\bibitem[\protect\citeauthoryear{Kralj-Iglič et~al.}{2002}]{Kralj2002}
\begin{barticle}
\bauthor{\bsnm{Kralj-Iglič}, \binits{V.}},
\bauthor{\bsnm{Remškar}, \binits{M.}},
\bauthor{\bsnm{Vidmar}, \binits{G.}},
\bauthor{\bsnm{Fošnarič}, \binits{M.}},
\bauthor{\bsnm{Iglič}, \binits{A.}}:
\batitle{Deviatoric elasticity as a possible physical mechanism explaining collapse of inorganic micro and nanotubes}.
\bjtitle{Physics Letters A}
\bvolume{296}(\bissue{2}),
\bfpage{151}--\blpage{155}
(\byear{2002})
\end{barticle}
\endbibitem

\bibitem[\protect\citeauthoryear{Enyashin et~al.}{2007}]{Enyashin2007}
\begin{barticle}
\bauthor{\bsnm{Enyashin}, \binits{A.}},
\bauthor{\bsnm{Gemming}, \binits{S.}},
\bauthor{\bsnm{Seifert}, \binits{G.}}:
\batitle{Nanosized allotropes of molybdenum disulfide}.
\bjtitle{The European Physical Journal Special Topics}
\bvolume{149}(\bissue{1}),
\bfpage{103}--\blpage{125}
(\byear{2007})
\end{barticle}
\endbibitem

\bibitem[\protect\citeauthoryear{Remskar et~al.}{2022}]{Remskar2022}
\begin{barticle}
\bauthor{\bsnm{Remskar}, \binits{M.}},
\bauthor{\bsnm{Hüttel}, \binits{A.K.}},
\bauthor{\bsnm{Shubina}, \binits{T.V.}},
\bauthor{\bsnm{Seabaugh}, \binits{A.}},
\bauthor{\bsnm{Fathipour}, \binits{S.}},
\bauthor{\bsnm{Lawrowski}, \binits{R.}},
\bauthor{\bsnm{Schreiner}, \binits{R.}}:
\batitle{Confinement related phenomena in {M}o{S}$_2$ tubular structures grown from vapour phase}.
\bjtitle{Israel Journal of Chemistry}
\bvolume{62}(\bissue{3-4}),
\bfpage{202100100}
(\byear{2022})
\end{barticle}
\endbibitem

\bibitem[\protect\citeauthoryear{Selig et~al.}{2016}]{Selig2016}
\begin{barticle}
\bauthor{\bsnm{Selig}, \binits{M.}},
\bauthor{\bsnm{Bergh{\"a}user}, \binits{G.}},
\bauthor{\bsnm{Raja}, \binits{A.}},
\bauthor{\bsnm{Nagler}, \binits{P.}},
\bauthor{\bsnm{Sch{\"u}ller}, \binits{C.}},
\bauthor{\bsnm{Heinz}, \binits{T.F.}},
\bauthor{\bsnm{Korn}, \binits{T.}},
\bauthor{\bsnm{Chernikov}, \binits{A.}},
\bauthor{\bsnm{Malic}, \binits{E.}},
\bauthor{\bsnm{Knorr}, \binits{A.}}:
\batitle{Excitonic linewidth and coherence lifetime in monolayer transition metal dichalcogenides}.
\bjtitle{Nature Communications}
\bvolume{7}(\bissue{1}),
\bfpage{13279}
(\byear{2016})
\end{barticle}
\endbibitem

\bibitem[\protect\citeauthoryear{Shubina et~al.}{2019}]{Shubina2019}
\begin{barticle}
\bauthor{\bsnm{Shubina}, \binits{T.V.}},
\bauthor{\bsnm{Remškar}, \binits{M.}},
\bauthor{\bsnm{Davydov}, \binits{V.Y.}},
\bauthor{\bsnm{Belyaev}, \binits{K.G.}},
\bauthor{\bsnm{Toropov}, \binits{A.A.}},
\bauthor{\bsnm{Gil}, \binits{B.}}:
\batitle{Excitonic emission in van der {W}aals nanotubes of transition metal dichalcogenides}.
\bjtitle{Annalen der Physik}
\bvolume{531}(\bissue{6}),
\bfpage{1800415}
(\byear{2019})
\end{barticle}
\endbibitem

\bibitem[\protect\citeauthoryear{Cadiz et~al.}{2017}]{Cadiz2017}
\begin{barticle}
\bauthor{\bsnm{Cadiz}, \binits{F.}},
\bauthor{\bsnm{Courtade}, \binits{E.}},
\bauthor{\bsnm{Robert}, \binits{C.}},
\bauthor{\bsnm{Wang}, \binits{G.}},
\bauthor{\bsnm{Shen}, \binits{Y.}},
\bauthor{\bsnm{Cai}, \binits{H.}},
\bauthor{\bsnm{Taniguchi}, \binits{T.}},
\bauthor{\bsnm{Watanabe}, \binits{K.}},
\bauthor{\bsnm{Carrere}, \binits{H.}},
\bauthor{\bsnm{Lagarde}, \binits{D.}},
\bauthor{\bsnm{Manca}, \binits{M.}},
\bauthor{\bsnm{Amand}, \binits{T.}},
\bauthor{\bsnm{Renucci}, \binits{P.}},
\bauthor{\bsnm{Tongay}, \binits{S.}},
\bauthor{\bsnm{Marie}, \binits{X.}},
\bauthor{\bsnm{Urbaszek}, \binits{B.}}:
\batitle{Excitonic linewidth approaching the homogeneous limit in {M}o{S}$_2$-based van der {W}aals heterostructures}.
\bjtitle{Phys. Rev. X}
\bvolume{7},
\bfpage{021026}
(\byear{2017})
\end{barticle}
\endbibitem

\bibitem[\protect\citeauthoryear{Musfeldt et~al.}{2020}]{Musfeldt2020}
\begin{barticle}
\bauthor{\bsnm{Musfeldt}, \binits{J.L.}},
\bauthor{\bsnm{Iwasa}, \binits{Y.}},
\bauthor{\bsnm{Tenne}, \binits{R.}}:
\batitle{{Nanotubes from layered transition metal dichalcogenides}}.
\bjtitle{Physics Today}
\bvolume{73}(\bissue{8}),
\bfpage{42}--\blpage{48}
(\byear{2020})
\end{barticle}
\endbibitem

\bibitem[\protect\citeauthoryear{Kazanov et~al.}{2018}]{Kazanov2018}
\begin{barticle}
\bauthor{\bsnm{Kazanov}, \binits{D.R.}},
\bauthor{\bsnm{Poshakinskiy}, \binits{A.V.}},
\bauthor{\bsnm{Davydov}, \binits{V.Y.}},
\bauthor{\bsnm{Smirnov}, \binits{A.N.}},
\bauthor{\bsnm{Eliseyev}, \binits{I.A.}},
\bauthor{\bsnm{Kirilenko}, \binits{D.A.}},
\bauthor{\bsnm{Remškar}, \binits{M.}},
\bauthor{\bsnm{Fathipour}, \binits{S.}},
\bauthor{\bsnm{Mintairov}, \binits{A.}},
\bauthor{\bsnm{Seabaugh}, \binits{A.}},
\bauthor{\bsnm{Gil}, \binits{B.}},
\bauthor{\bsnm{Shubina}, \binits{T.V.}}:
\batitle{{Multiwall {M}o{S}$_2$ tubes as optical resonators}}.
\bjtitle{Applied Physics Letters}
\bvolume{113}(\bissue{10}),
\bfpage{101106}
(\byear{2018})
\end{barticle}
\endbibitem

\bibitem[\protect\citeauthoryear{Sun et~al.}{2008}]{Sun2008}
\begin{barticle}
\bauthor{\bsnm{Sun}, \binits{L.}},
\bauthor{\bsnm{Chen}, \binits{Z.}},
\bauthor{\bsnm{Ren}, \binits{Q.}},
\bauthor{\bsnm{Yu}, \binits{K.}},
\bauthor{\bsnm{Bai}, \binits{L.}},
\bauthor{\bsnm{Zhou}, \binits{W.}},
\bauthor{\bsnm{Xiong}, \binits{H.}},
\bauthor{\bsnm{Zhu}, \binits{Z.Q.}},
\bauthor{\bsnm{Shen}, \binits{X.}}:
\batitle{Direct observation of whispering gallery mode polaritons and their dispersion in a zno tapered microcavity}.
\bjtitle{Phys. Rev. Lett.}
\bvolume{100},
\bfpage{156403}
(\byear{2008})
\end{barticle}
\endbibitem

\bibitem[\protect\citeauthoryear{Kazanov et~al.}{2020}]{Kazanov2020}
\begin{botherref}
\oauthor{\bsnm{Kazanov}, \binits{D.}},
\oauthor{\bsnm{Rakhlin}, \binits{M.}},
\oauthor{\bsnm{Poshakinskiy}, \binits{A.}},
\oauthor{\bsnm{Shubina}, \binits{T.}}:
Towards exciton-polaritons in an individual {M}o{S}$_2$ nanotube.
Nanomaterials
\textbf{10}(2)
(2020)
\end{botherref}
\endbibitem

\bibitem[\protect\citeauthoryear{Yadgarov et~al.}{2018}]{Yadgarov2018}
\begin{barticle}
\bauthor{\bsnm{Yadgarov}, \binits{L.}},
\bauthor{\bsnm{Višić}, \binits{B.}},
\bauthor{\bsnm{Abir}, \binits{T.}},
\bauthor{\bsnm{Tenne}, \binits{R.}},
\bauthor{\bsnm{Polyakov}, \binits{A.Y.}},
\bauthor{\bsnm{Levi}, \binits{R.}},
\bauthor{\bsnm{Dolgova}, \binits{T.V.}},
\bauthor{\bsnm{Zubyuk}, \binits{V.V.}},
\bauthor{\bsnm{Fedyanin}, \binits{A.A.}},
\bauthor{\bsnm{Goodilin}, \binits{E.A.}},
\bauthor{\bsnm{Ellenbogen}, \binits{T.}},
\bauthor{\bsnm{Tenne}, \binits{R.}},
\bauthor{\bsnm{Oron}, \binits{D.}}:
\batitle{Strong light–matter interaction in tungsten disulfide nanotubes}.
\bjtitle{Phys. Chem. Chem. Phys.}
\bvolume{20},
\bfpage{20812}--\blpage{20820}
(\byear{2018})
\end{barticle}
\endbibitem

\bibitem[\protect\citeauthoryear{Graf et~al.}{2016}]{Graf2016}
\begin{barticle}
\bauthor{\bsnm{Graf}, \binits{A.}},
\bauthor{\bsnm{Tropf}, \binits{L.}},
\bauthor{\bsnm{Zakharko}, \binits{Y.}},
\bauthor{\bsnm{Zaumseil}, \binits{J.}},
\bauthor{\bsnm{Gather}, \binits{M.C.}}:
\batitle{Near-infrared exciton-polaritons in strongly coupled single-walled carbon nanotube microcavities}.
\bjtitle{Nature Communications}
\bvolume{7}(\bissue{1}),
\bfpage{13078}
(\byear{2016})
\end{barticle}
\endbibitem

\bibitem[\protect\citeauthoryear{Sinha et~al.}{2020}]{Sinha2020}
\begin{barticle}
\bauthor{\bsnm{Sinha}, \binits{S.S.}},
\bauthor{\bsnm{Zak}, \binits{A.}},
\bauthor{\bsnm{Rosentsveig}, \binits{R.}},
\bauthor{\bsnm{Pinkas}, \binits{I.}},
\bauthor{\bsnm{Tenne}, \binits{R.}},
\bauthor{\bsnm{Yadgarov}, \binits{L.}}:
\batitle{Size-dependent control of exciton–polariton interactions in {W}{S}$_2$ nanotubes}.
\bjtitle{Small}
\bvolume{16}(\bissue{4}),
\bfpage{1904390}
(\byear{2020})
\end{barticle}
\endbibitem

\bibitem[\protect\citeauthoryear{Eliseyev et~al.}{2023}]{Eliseyev2023}
\begin{botherref}
\oauthor{\bsnm{Eliseyev}, \binits{I.A.}},
\oauthor{\bsnm{Borodin}, \binits{B.R.}},
\oauthor{\bsnm{Kazanov}, \binits{D.R.}},
\oauthor{\bsnm{Poshakinskiy}, \binits{A.V.}},
\oauthor{\bsnm{Remškar}, \binits{M.}},
\oauthor{\bsnm{Pavlov}, \binits{S.I.}},
\oauthor{\bsnm{Kotova}, \binits{L.V.}},
\oauthor{\bsnm{Alekseev}, \binits{P.A.}},
\oauthor{\bsnm{Platonov}, \binits{A.V.}},
\oauthor{\bsnm{Davydov}, \binits{V.Y.}},
\oauthor{\bsnm{Shubina}, \binits{T.V.}}:
Twisted nanotubes of transition metal dichalcogenides with split optical modes for tunable radiated light resonators.
Advanced Optical Materials,
2202782
(2023)
\end{botherref}
\endbibitem

\bibitem[\protect\citeauthoryear{Landau and Lifshits}{1991}]{Landau3}
\begin{bbook}
\bauthor{\bsnm{Landau}, \binits{L.D.}},
\bauthor{\bsnm{Lifshits}, \binits{E.M.}}:
\bbtitle{Quantum Mechanics: Non-Relativistic Theory}.
\bsertitle{Course of theoretical physics},
vol. \bseriesno{3}.
\bpublisher{Elsevier Science}, \blocation{???}
(\byear{1991})
\end{bbook}
\endbibitem

\bibitem[\protect\citeauthoryear{Ghorbani-Asl et~al.}{2013}]{Ghorbani-Asl2013}
\begin{barticle}
\bauthor{\bsnm{Ghorbani-Asl}, \binits{M.}},
\bauthor{\bsnm{Zibouche}, \binits{N.}},
\bauthor{\bsnm{Wahiduzzaman}, \binits{M.}},
\bauthor{\bsnm{Oliveira}, \binits{A.F.}},
\bauthor{\bsnm{Kuc}, \binits{A.}},
\bauthor{\bsnm{Heine}, \binits{T.}}:
\batitle{Electromechanics in {M}o{S}$_2$ and {W}{S}$_2$: nanotubes vs. monolayers}.
\bjtitle{Scientific Reports}
\bvolume{3}(\bissue{1}),
\bfpage{2961}
(\byear{2013})
\end{barticle}
\endbibitem

\bibitem[\protect\citeauthoryear{Peelaers and Van~de Walle}{2012}]{Peelaers2012}
\begin{barticle}
\bauthor{\bsnm{Peelaers}, \binits{H.}},
\bauthor{\bsnm{Walle}, \binits{C.G.}}:
\batitle{Effects of strain on band structure and effective masses in {M}o{S}$_2$}.
\bjtitle{Phys. Rev. B}
\bvolume{86},
\bfpage{241401}
(\byear{2012})
\end{barticle}
\endbibitem

\bibitem[\protect\citeauthoryear{Conley et~al.}{2013}]{Conley2013}
\begin{barticle}
\bauthor{\bsnm{Conley}, \binits{H.J.}},
\bauthor{\bsnm{Wang}, \binits{B.}},
\bauthor{\bsnm{Ziegler}, \binits{J.I.}},
\bauthor{\bsnm{Haglund~Jr.}, \binits{R.F.}},
\bauthor{\bsnm{Pantelides}, \binits{S.T.}},
\bauthor{\bsnm{Bolotin}, \binits{K.I.}}:
\batitle{Bandgap engineering of strained monolayer and bilayer {M}o{S}$_2$}.
\bjtitle{Nano Letters}
\bvolume{13}(\bissue{8}),
\bfpage{3626}--\blpage{3630}
(\byear{2013})
\end{barticle}
\endbibitem

\bibitem[\protect\citeauthoryear{Lloyd et~al.}{2016}]{lloyd2016}
\begin{barticle}
\bauthor{\bsnm{Lloyd}, \binits{D.}},
\bauthor{\bsnm{Liu}, \binits{X.}},
\bauthor{\bsnm{Christopher}, \binits{J.W.}},
\bauthor{\bsnm{Cantley}, \binits{L.}},
\bauthor{\bsnm{Wadehra}, \binits{A.}},
\bauthor{\bsnm{Kim}, \binits{B.L.}},
\bauthor{\bsnm{Goldberg}, \binits{B.B.}},
\bauthor{\bsnm{Swan}, \binits{A.K.}},
\bauthor{\bsnm{Bunch}, \binits{J.S.}}:
\batitle{Band gap engineering with ultralarge biaxial strains in suspended monolayer {M}o{S}$_2$}.
\bjtitle{Nano Lett.}
\bvolume{16}(\bissue{9}),
\bfpage{5836}--\blpage{5841}
(\byear{2016})
\end{barticle}
\endbibitem

\bibitem[\protect\citeauthoryear{Alpeggiani et~al.}{2018}]{Alpeggiani2018}
\begin{barticle}
\bauthor{\bsnm{Alpeggiani}, \binits{F.}},
\bauthor{\bsnm{Gong}, \binits{S.-H.}},
\bauthor{\bsnm{L.}, \binits{K.}}:
\batitle{Dispersion and decay rate of exciton-polaritons and radiative modes in transition metal dichalcogenide monolayers}.
\bjtitle{Phys. Rev. B}
\bvolume{97},
\bfpage{205436}
(\byear{2018})
\end{barticle}
\endbibitem

\bibitem[\protect\citeauthoryear{Wang et~al.}{2016}]{Wang2016}
\begin{barticle}
\bauthor{\bsnm{Wang}, \binits{Q.}},
\bauthor{\bsnm{Sun}, \binits{L.}},
\bauthor{\bsnm{Zhang}, \binits{B.}},
\bauthor{\bsnm{Chen}, \binits{C.}},
\bauthor{\bsnm{Shen}, \binits{X.}},
\bauthor{\bsnm{W.}, \binits{L.}}:
\batitle{Direct observation of strong light-exciton coupling in thin ws2 flakes}.
\bjtitle{Opt. Express}
\bvolume{24},
\bfpage{7151}
(\byear{2016})
\end{barticle}
\endbibitem

\bibitem[\protect\citeauthoryear{Lundt et~al.}{2017}]{Lundt2017}
\begin{barticle}
\bauthor{\bsnm{Lundt}, \binits{N.}},
\bauthor{\bsnm{Stoll}, \binits{N.S.}},
\bauthor{\bsnm{Nagler}, \binits{P.}},
\bauthor{\bsnm{Nalitov}, \binits{A.}},
\bauthor{\bsnm{Klembt}, \binits{S.}},
\bauthor{\bsnm{Betzold}, \binits{S.}},
\bauthor{\bsnm{Goddard}, \binits{J.}},
\bauthor{\bsnm{Frieling}, \binits{E.}},
\bauthor{\bsnm{Kavokin}, \binits{A.V.}},
\bauthor{\bsnm{Schüller}, \binits{C.}},
\bauthor{\bsnm{Korn}, \binits{T.}},
\bauthor{\bsnm{Höfling}, \binits{S.}},
\bauthor{\bsnm{Schneider}, \binits{C.}}:
\batitle{Observation of macroscopic valley-polarized monolayer exciton-polaritons at room temperature}.
\bjtitle{Phys. Rev. B}
\bvolume{96},
\bfpage{241403}
(\byear{2017})
\end{barticle}
\endbibitem

\end{thebibliography}

\end{document}


\title[Article Title]{Supplementary materials for article "Direct observation of split-mode exciton-polaritons in a single MoS$_2$ nanotube" 

}


\author[1]{\fnm{A.I.} \sur{Galimov}}
\author[1]{\fnm{D.R.} \sur{Kazanov}}
\author[1]{\fnm{A.V.} \sur{Poshakinskiy}}
\author[1]{\fnm{M.V.} \sur{Rakhlin}}
\author[1]{\fnm{I.A.} \sur{Eliseyev}}
\author[1]{\fnm{A.A.} \sur{Toropov}}
\author[2]{\fnm{M.} \sur{Remskar}}
\author*[1]{\fnm{T.V.} \sur{Shubina}\email{shubina@beam.ioffe.ru}}






\maketitle

\section*{An optical double-well potential}

\begin{figure}[t]
    \centering
    \includegraphics[width=0.99\columnwidth]{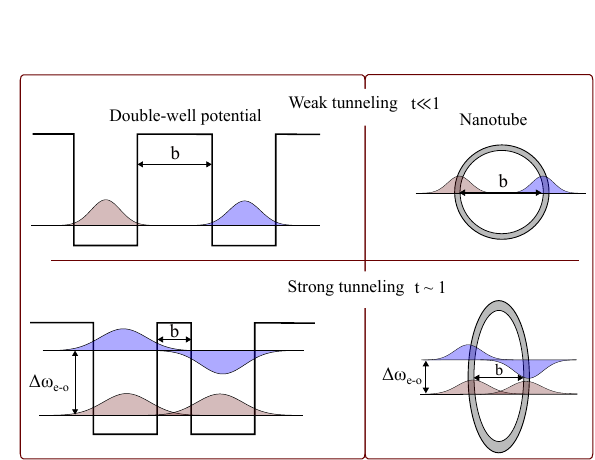}
    \caption{A similarity between a particle in a double-well potential and an optical mode in a NT wall. 
    (Upper part) No tunneling with thick barrier (gap), wave functions (electric fields) do not overlap, degenerate eigenenergies.
    (Lower part) Strong tunneling with thin barrier (gap), eigenstates are mixed, and the eigenenergies are split}
    \label{fig:Wavefunction2_1}
\end{figure}

There is an analogy between a double-well potential for a particle and that for electric field in a NT with a flattened cross-section (Fig.~\ref{fig:Wavefunction2_1}). If the thickness of the barrier between the two potential wells  $d$ is very large, the particles can reside in the left or right well, and the eigenstates are twice degenerate. When the thickness of the barrier is decreased, the tunneling between the wells arises and mixes the wave functions of the particle states. Then, the eigenstates are the even and odd superposition of the states in the two wells, i.e. $\Psi_{\rm{e,o}}= \frac{1}{\sqrt{2}}(\psi_1\mp \psi_2) $ and the corresponding eigenenergies are equal to $E_{\rm{e,o}} = E_0(1 \pm t)$. The magnitude of the splitting is determined by the exponential factor $t \propto e^{-\kappa d}$, where $\kappa$ is the imaginary part of the wave vector of the particle in the barrier, which grows rapidly as the wells move closer to each other.

A similar phenomenon occurs with electromagnetic field in a NT. In the non-deformed NT, the field is strongly confined inside the wall and the tunneling is negligible. The eigenstates, characterized by their azimuthal number $m$, can propagate clockwise or counterclockwise. Thus, they are twice-degenerate, since their frequency $\omega_m = mc/(r n_{\rm eff})$ depends only on the $m$ and NT radius $r$. Here $c$ is the speed of light and $n_{\rm eff}$ is the effective refractive index of the mode, which is some average between the refractive index of the NT material and air.
As we squeeze the NT cross-section, the gap between the longer parts decreases and the evanescent tails begin to overlap. Now the eigen states represent the standing waves, even and odd with respect to major axis of the cross-section. The energy of the split modes are $\hbar\omega_{\rm{e,o}}=\hbar\omega_0(1 \mp t_{\rm{WGM}})$. With further deformation of the cross-section, the splitting increases exponentially $t_{\rm{WGM}}\propto e^{-\kappa_r b}$, where $b$ is the minor axis of the NT cross-section and $\kappa_r=\omega/c\sqrt{n^2_{\rm{eff}}-1}$ is the decay rate of the evanescent tail outside the NT wall. In the limit of a fully collapsed NT, the system reduces to a conventional Fabri-P\'erot resonator. 

\section*{NT characterization}

\begin{figure}[h]
    \centering
    \includegraphics[width=1\columnwidth]{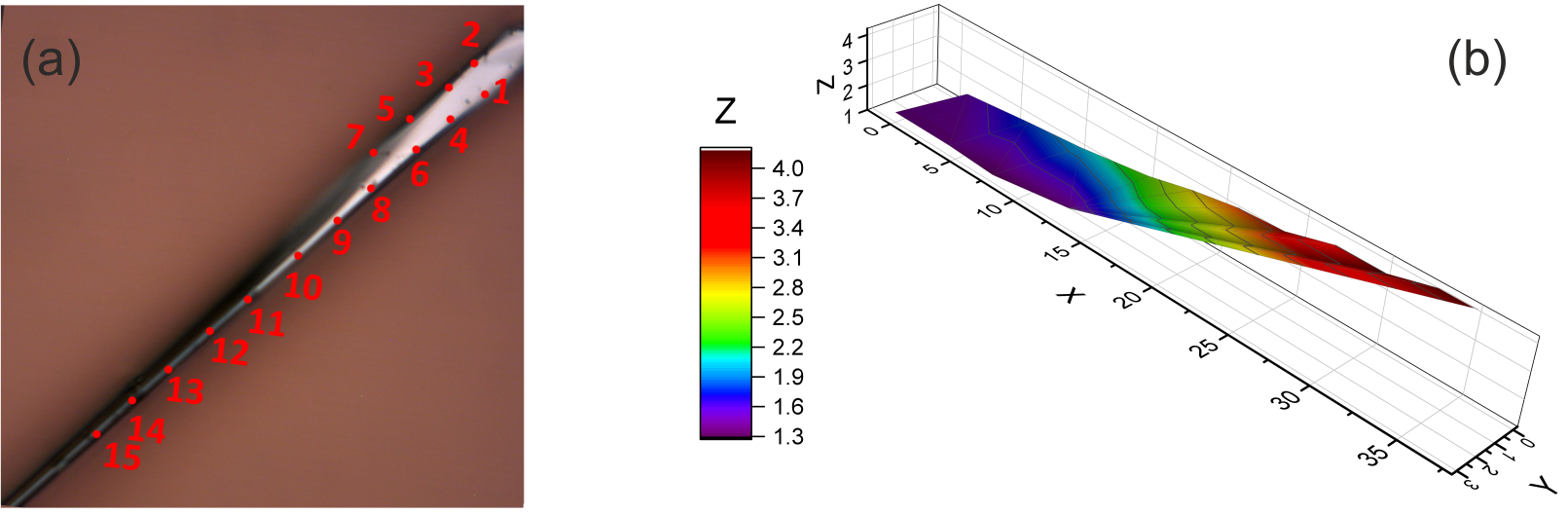}
    \caption{(a) Bright-field image of the studied NT. The numbers indicate points where the NT height was measured. (b) Measured NT profile. Distances and height are given in micrometers. $(X,Y)= (0,0)$ corresponds to point No. 1 in (a).}
    \label{fig:Tube_b}
\end{figure}

To determine the tube shape, we carried out optical measurements using a microscope in the Horiba LabRam HR Evolution spectrometer. The microscope is equipped with a precision XY stage SCANPlus and a motorized focus drive MA-42 (M{\"a}rzh{\"a}user, Wetzlar, Germany), which allowed us to control the focusing distance (Z coordinate) and the position in the substrate plane (X and Y coordinates) with high accuracy ($\sim 20$ nm). 

The focal point and, therefore, the Z coordinate was determined by the maximum reflection when illuminated with white light. In Fig. \ref{fig:Tube_b}(b) shows the shape and dimensions of the tube obtained from the 15 points marked in Fig. \ref{fig:Tube_b}(a). Here the zero point in the Z direction corresponds to the substrate plane. From this experiment we can estimate that the radius of the undeformed equivalent tube is $\sim 800-1000$ nm, the aspect ratio of the tube at points No. 1, 2 is about 1:3, and at point No. 15 it is about 1:10.

\section*{Raman measurements}
\begin{figure}[t]
    \centering
    \includegraphics[width=0.8\columnwidth]{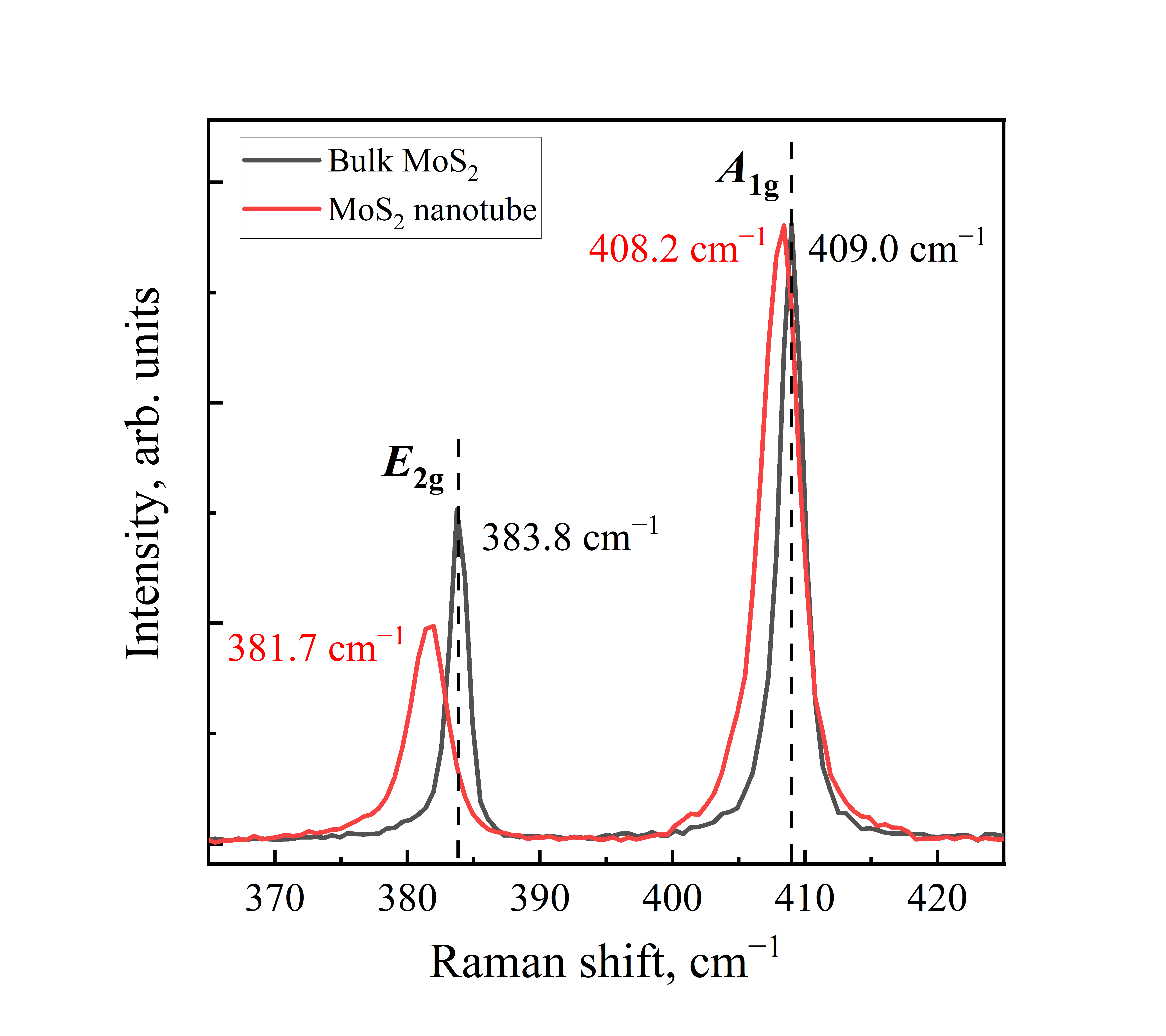}
    \caption{Raman spectra of the NT and a bulk MoS$_2$ crystal. Dashed lines demonstrate the positions of $\rm{E}_{\rm{2g}}$ and $\rm{A}_{\rm{1g}}$ lines of bulk MoS$_2$. Frequencies of the $\rm{E}_{\rm{2g}}$ and $\rm{A}_{\rm{1g}}$ lines are indicated for each of the lines in the corresponding color.}
    \label{fig:Raman}
\end{figure}
First, it is worth noting that the effects of deformation differ in single-walled and multi-walled NTs. All experimental samples are multi-walled, and single-walled NTs are realized very rarely \cite{Remskar2001}, but are a favorite object of theoretical research. \cite{Seifert2000a, Zibouche2012, Wu2018, Hisama2021}. In single-walled NTs, the tensile strain due to surface curvature is significant at small diameters, but as the diameter increases, the strain decreases because the flatter the surface, the smaller the stretching of atomic bonds. As shown by Seifert et al. \cite{Seifert2000a, Seifert2000b}, this is accompanied by an increase in the band gap energy (blue shift) up to the planar monolayer limit. A different situation is realized in multi-walled NTs, when each subsequent monolayer must maintain the stacking order with respect to the previous layers, stretching over them. The resulting tensile strain reduces the band gap \cite{Ghorbani-Asl2013}, similar to what is observed in atomically-thin layers with tension \cite{Peelaers2012a, Conley2013}. 
In additon, the energy gap between excitons A and B may deviate from the average due to the difference in the change in their energies upon deformation \cite{lloyd2016}.

To show that the studied NT undergoes tensile deformation, we performed combined micro-Raman and micro-PL studies.
When comparing the Raman spectrum of a NT with the spectrum of freshly exfoliated bulk MoS$_2$ (Fig.~\ref{fig:Raman}), we observe a significant redshift of both $\rm{E}_{\rm{2g}}$ and $\rm{A}_{\rm{1g}}$ lines. In general, the redshift of Raman lines indicates the presence of tensile strain, which is expected to be uniaxial in the tubes. 
Quantitative estimation of the magnitude of strain in a multiwall NT based on the Raman data is a complex task. According to the data of several existing works on uniaxially strained multilayer MoS$_2$ 
\cite{yang2014, Tan2018a}, such a shift may point to the strain values from  $\sim (1.5 - 2) \%$ \cite{yang2014} to $\sim (3 - 8) \%$ \cite{Tan2018a}. According to the strain-induced direct band gap shift rate obtained in \cite{yang2014}, the observed A-exciton shift ($\sim 200$ meV) corresponds to $\sim 4 \%$ tensile strain. However, these estimates are qualitative in nature, since the deformation in the NT is very specific and possible 3R stacking in the NT layers may contribute to this shift. Only the tensile strain $\geq 1 \%$ can be reliably indicated, which justifies the red shift of the exciton resonances.

\section*{Model of reflection from a flattened NT}
Let the plane wave $E_{\rm inc}=e^{\rm{i}\bf{k r}}$ polarized along the NT axis fall on the NT in normal incidence at angle $\alpha$ with the minor axis.The matrix element $t^{(m)}_{\rm in}$ of excitation of a  even or odd WGM $E^{\rm (e,o)}_m$ is determined by the overlap integral,
\begin{equation}
t^{(m)}_{\rm in}=\int{E^{\rm (e,o)}_m E_{\rm inc}~d\varphi}
\end{equation}
as in \cite{Eliseyev2023}. Accordingly, the electric field amplitude inside the NT is expressed as 
\begin{equation}
A^{(m)}_{\rm mode}=\frac{t^{(m)}_{\rm in}}{\omega-\omega^{(m)}_{\rm e,o}(b)+{\rm i}(\Gamma + \Gamma_0)},
\end{equation}
where $\omega^{(m)}_{\rm e,o}(b) \approx \omega^{(m)}_0 (1\mp C e^{-(\omega/c) b \sqrt{n^2_{\rm eff}-1} })$ is the dispersion of even/odd WGMs with angular number $m$, $\Gamma, \Gamma_0$ are homogeneous and inhomogeneous broadening of WGMs, respectively. 

This is followed by the reverse process of light emission from the NT. The efficiency of the scattering process is also determined by the overlap of the WGM electric field and the scattered electric field
\begin{equation}
t^{*(m)}_{\rm out}=\int{E^{\rm (e,o)}_m E^*_{\rm scatt}~d\varphi}
\end{equation}
so the scattered field amplitude reads
\begin{equation}
A^{(m)}_{\rm scat}=A^{(m)}_{\rm mode} t^{*(m)}_{\rm out}
\end{equation}
Thus, full reflection $R$ would be a sum of all modes that interact with incident field 
\begin{equation}\label{eq:R}
R(\omega, b) \propto \bigg |\sum_{m} A^{(m)}_{\rm scat}\bigg |^2 = \bigg |\sum_{m} \frac{t^{*(m)}_{\rm out} t^{(m)}_{\rm in}}{\omega-\omega^{(m)}_0  (1\mp C e^{-(\omega/c) b \sqrt{n^2_{\rm eff}-1} }) +{\rm i}(\Gamma + \Gamma_0)}\bigg |^2
\end{equation}

For weakly deformed NTs, the matrix element of the WGM excitation can be found as a convolution of the WGM electric field $E_{\rm mode} = E_0 {\rm{sin}}(m\varphi)$ or $E_0 {\rm{cos}}(m\varphi)$  with incident electric field  in a form of a plane wave $E_{\rm inc} =  \tilde{E_0}e^{{\rm{i}} k_z r {\rm{sin}}(\varphi-\alpha)} $ propagating in the direction determined by the angle $\alpha$. Following the decomposition of a plane wave in cylindrical waves $e^{{\rm i} z {\rm sin}\varphi} = \sum_{n=-\infty}^{\infty} J_n(z) e^{{\rm i}n\varphi}$ we express the coupling efficiency, for example, for odd mode as
\begin{align}
    t^{(m)}_{\rm in} = \int{E^{(e)}_{\rm{m}}E_{{\rm inc}} d\varphi} = \int{E_0 {\rm sin}(m\varphi)\tilde{E}_0 e^{{\rm i} k_z r sin (\varphi - \alpha)} d\varphi} = \\ \sum^{+\infty}_{n=-\infty} E_0 \tilde{E}_0 J_n (k_z r) \int{{\rm sin}(m\varphi) e^{{\rm i} n (\varphi - \alpha)} d\varphi}.
\end{align}
Using properties of trigonometric functions one could obtain coupling efficiency for odd and even mode, respectively
\begin{align}\label{eq:coupling}
    t^{(o,m)}_{\rm in} = 2\sqrt{2} \pi E_0 \tilde{E}_0 e^{{\rm i} \frac{\pi}{4}(2m-1)}J_m(k_z r) {\rm sin}(m[\alpha + \frac{\pi}{2}]) \\
    t^{(e,m)}_{\rm in} = 2\sqrt{2} \pi E_0 \tilde{E}_0 e^{{\rm i} \frac{\pi}{4}(2m-1)}J_m(k_z r) {\rm cos}(m[\alpha + \frac{\pi}{2}]).
\end{align}

\begin{figure}[b]
    \centering
    \includegraphics[width=0.99\columnwidth]{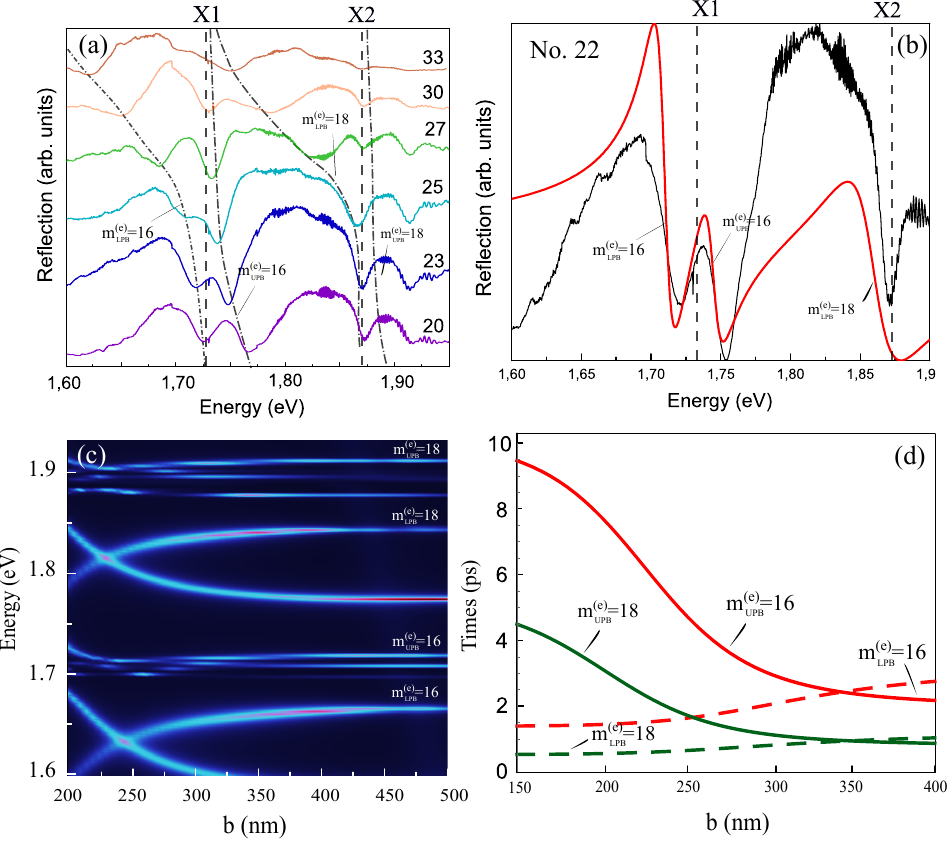}
    \caption{ (a) Raw reflection spectra selected among the region of points 20---33 of Fig. 2 from main text. Thin lines show upper and lower polariton branches of two optical modes with $m=16, 18$. (b) Experimental reflection spectra (black line) corresponding to the point 22 and its fit (red line) by eq. (\ref{eq:R}) for several WGMs near A- and B- exciton resonances. (c) Calculated optical modes for studied NT depending on the degree of flattening. (d) Decay times of two optical modes with $m=16, 18$ for UPB and LPB. Solid line and dashed line are UPB and LPB for modes $m=16, 18$, respectfully.}
    \label{fig:R_analytical}
\end{figure}

\section*{Determination of optical parameters }

Figure \ref{fig:R_analytical} illustrates the process of determining the optical parameters of NT. The raw reflectance spectra are shown in Fig. \ref{fig:R_analytical}a. We will focus on two modes close to exciton resonances with angular numbers $m=16$ and $m=18$. The dispersion of WGM polaritons is shown by thin dash-dot lines. To obtain WGM mode energies near exciton resonances, we fit the experimental spectra with analytical model for reflection Eq. (\ref{eq:R}) (red line in Fig. \ref{fig:R_analytical}b). The decay time obtained from the analytical approximation is about $0.5-3$ ps for both modes. Also, the same results (Fig. \ref{fig:R_analytical}c) were obtained from numerical calculations of WGM for the studied NTs with flattening. Using the decay rates of optical modes and the radiative decay rates of A- and B-exciton resonances ($\sim 5$ ps) allows us to estimate the hybrid exciton-polariton decay times. Figure \ref{fig:R_analytical}(c) shows its behavior of several WGMs obtained by solving Maxwell's equations. The inverse imaginary part of the solution expresses the decay times of the UPB and LPB of polaritons with varying degrees of flattening of the NT cross-section
(Fig. \ref{fig:R_analytical} (d)). 
Strongly flattened tube ($b=150$ nm) shows a regime of pure optical modes for LPBs and decay times close to pure decay rates of WGMs, $0.5-2$ ps. And UPBs have decay rates of almost pure excitons. The corresponding slowdown in recombination rate means the enhanced interaction of the upper branch with exciton resonances. Here we took decay rates of 5 and 10 ps to clearly show the behavior of the polariton branches.
Another case is a weakly flattened NT ($b>350$ nm), when we observe modified decay times for both UPB and LPB, provided that optical modes and exciton resonances are close.
